\documentclass[conference,compsoc]{IEEEtran}

\usepackage{times}
\usepackage{algorithmic}
\usepackage{amsmath}
\usepackage{amssymb}
\usepackage{color}
\usepackage{csquotes}
\usepackage{graphicx}
\usepackage{mathrsfs}
\usepackage{textcomp}
\usepackage{multirow}
\usepackage{url}
\usepackage{tikz}
\usepackage{listings}
\usepackage{xspace}
\usepackage{float}
 
\def \VersionWithComments {}

\ifdefined \VersionWithComments
	\usepackage{marginnote}
	\newcommand{\marginX}{\marginnote{\huge{\quad\quad\textbf{!}\quad\quad}}}

	\newcommand{\instructions}[1]{{\color{red}\marginX{}\textbf{[Instructions: ``#1'']}}}
	\newcommand{\reviewer}[2]{\mbox{}{\color{red}\marginX{}\textbf{[Reviewer #1}: ``#2'']}}
 	\newcommand{\todo}[1]{\mbox{}{\color{red}{\marginX{}\textbf{TODO}\ifx#1\\\else:\ \fi #1}}}
\else
	\newcommand{\instructions}[1]{}
	\newcommand{\ea}[1]{}
	\newcommand{\rh}[1]{}
	\newcommand{\ohr}[1]{}
        \newcommand{\sq}[1]{}
        \newcommand{\sg}[1]{}
	\newcommand{\reviewer}[2]{}
 	\newcommand{\todo}[1]{}

\fi

\definecolor{mygreen}{rgb}{0,0.6,0}
\newcommand{\New}[1]{#1}
\newcommand{\old}[1]{}

\newcommand{\ignore}[1]{}  %

\newtheorem{theorem}{Theorem}

\newtheorem{definition}{Definition}

\newtheorem{proposition}{Proposition}
\newtheorem{example}{Example}

\newcommand{\rbs}{RBS}

\newcommand{\code}[1]{\texttt{#1}}
\newcommand{\customer}{\mbox{\scshape{Mimos\xspace}}}

\newcommand{\pref}{\sqsubseteq} %
\newcommand{\Nod}{\mbox{\textbf n}} %

\newcommand{\Val}{\ensuremath{\mathbb{D}}}       %
\newcommand{\Time}{\ensuremath{\mathbb{R}}}      %
\newcommand{\Streams}{\ensuremath{\mathbb{S}}} %
\newcommand{\Timed}{\ensuremath{\mathbb{T}}}      %
\newcommand{\Str}{\mbox{\textbf S}} %
\newcommand{\TStr}{\mbox{\textbf T}} %
\newcommand{\sampTStr}{\TStr^{S}} %
\newcommand{\forget}[1]{}%

\newcommand{\conc}{\ensuremath{\cdot}} %
\newcommand{\rem}{\ensuremath{\mathbf{R}}} %
\newcommand{\Fun}{\ensuremath{\mathbf{F}}} %
\newcommand{\SysFun}{\ensuremath{\mathbf{F_s}}} %
\newcommand{\fun}{\ensuremath{\mathbf{f}}} %
\newcommand{\first}{\ensuremath{\mathbf{First}}} %
\newcommand{\mem}{\ensuremath{\mathbf{app}}} %
\newcommand{\ready}{\ensuremath{\mathbf{ready}}} %
\newcommand{\dready}{\ensuremath{\mathbf{d\_ready}}} %
\newcommand{\delay}{\ensuremath{\mathbf{delay}}} %
\newcommand{\deadl}{\ensuremath{\mathbf{w\_time}}} %
\newcommand{\readF}{\ensuremath{\mathbf{read}_{\Fun}}}
\newcommand{\writeF}{\ensuremath{\mathbf{write}_{\Fun}}}
\newcommand{\writes}{\ensuremath{\mathbf{write}}}
\newcommand{\outp}{\ensuremath{\mathbf{outp}}} %
\newcommand{\sens}{\ensuremath{\mathbf{rreg}}} %
\newcommand{\merge}{\ensuremath{\mathbf{merge}}} %

\newcommand{\etoe}{\ensuremath{\mathbf{e2e}}}

\usepackage{stackengine}

\usepackage{xcolor}

\definecolor{codegreen}{rgb}{0,0.6,0}
\definecolor{codegray}{rgb}{0.5,0.5,0.5}
\definecolor{codepurple}{rgb}{0.58,0,0.82}
\definecolor{backcolour}{rgb}{0.95,0.95,0.92}

\lstdefinestyle{mylststyle}{
    backgroundcolor=\color{white},   
    commentstyle=\color{codegreen},
    keywordstyle=\color{black},
    stringstyle=\color{codepurple},
    basicstyle=\ttfamily\scriptsize,
    breakatwhitespace=false,         
    breaklines=true,                 
    captionpos=b,                    
    keepspaces=true,                 
    showspaces=false,                
    showstringspaces=false,
    showtabs=false,                  
    tabsize=2,    
    frame=single
}

\newcommand\blfootnote[1]{%
  \begingroup
  \renewcommand\thefootnote{}\footnote{#1}%
  \addtocounter{footnote}{-1}%
  \endgroup
} 
\begin{document}

\title{
\customer$^*$
: A Deterministic Model
\\
for the
Design and Update of Real-Time Systems
}

\author{\IEEEauthorblockN{
Wang Yi\IEEEauthorrefmark{2},
Morteza Mohaqeqi\IEEEauthorrefmark{2} and
Susanne Graf\IEEEauthorrefmark{3}
}
\IEEEauthorblockA{\IEEEauthorrefmark{2}Uppsala University, Sweden\\ Email: \{wang.yi,morteza.mohaqeqi\}@it.uu.se}
\IEEEauthorblockA{\IEEEauthorrefmark{3}Univ. Grenoble Alpes, CNRS, Grenoble INP, VERIMAG, Grenoble, France\\
Email: susanne.graf@imag.fr}
}

\maketitle  

\begin{abstract}
Inspired by the pioneering work of Gilles Kahn on concurrent systems, we propose to model timed systems as a network of software components (implemented as real-time processes or tasks), each of which is specified to compute a collection of functions according to given timing constraints. We present a fixed-point semantics for this model which shows that each system function of such a network computes for a given set of (timed) input streams, a deterministic (timed) output stream. 
As a desired feature, such a network model can be modified by integrating new components for adding new system functions without changing the existing ones. Additionally, existing components may be replaced also by new ones fulfilling given requirements. Thanks to the deterministic semantics, a model-based approach is enabled for not only building systems but also updating them after deployment, allowing for efficient analysis techniques such as model-in-the-loop simulation to verify the complete behaviour of the updated system. \vspace{-1.5mm}
\end{abstract} \blfootnote{$^*$ \customer\  stands for \underline{M}ulti-\underline{I}nput \underline{M}ulti-\underline{O}utput Real-Time \underline{S}ystems.}
\section{Motivation}

Today, a large part of the functionality of technical systems such as cars, airplanes, and medical devices is implemented by software, as an (embedded) real-time system. The current trend is that traditionally mostly closed and single-purpose systems become open platforms. They aim at the integration of an expanding number of software components over their life-time, e.g., in order to customize and enhance their functionality according to varying needs of individual users. 
To enable this,
we must design and build systems that allow for updates after deployment. More importantly, it must be verified in field that the resulting systems not only preserve the original as well as the extended functionality, but also stay safe after updates.
Clearly, such analysis and verification processes must be carried out in a model-based approach. This demands a deterministic model for real-time systems, that supports for automatic synthesis of software components, and also ensures that all properties verified based on a system model (modified) are also true of the system (updated accordingly). %

Over the years, there have been a large number of timed models developed in the literature, notably the theory of timed automata, as well as various task models \cite{stigge2015} developed for real-time systems. These models are either extremely expressive and highly non-deterministic, but cannot be analyzed efficiently, or restrictive in terms of expressive power, and cannot be used to design systems (e.g., system with complex synchronization structures) with desired functional behaviours. 

In this paper, we propose to model timed systems as a network of real-time software components connected by communication channels in the style of Kahn Process Networks (KPN) \cite{kahn1974}, allowing asynchronous data exchange.
We present a simple but expressive description language, called \customer, 
to formalize such abstract network models.
We want to reduce the analysis complexity of functional and timing behaviours
of our model-based approach to system design and update. Therefore, we have
adopted the following principles for designing the model and its semantics:
\smallskip

{\bf Determinism.}
The key in a model-based approach is that the model of a system should be deterministic to ensure that the system behaves the same way as its model. 
In our approach, a model, as well as a system derived from it,
can be viewed as a stream transformer. 
For a given set of input streams, the output streams defined by the model (system) must be unique. This means that, it specifies a set of functions over streams (we call them \emph{system functions}) such that each one defines an output stream from a set of input streams. 
Second, the model (system) should be timing deterministic. This means that at any time point, if the inputs required for computation are available, 
the corresponding output should be delivered at a future time point after a fixed delay. Timing determinism can be relaxed to ensure only that the output may be delivered within a given time bound.

\smallskip

\forget{
{\bf Separation of functionality and timing.}
The model of a system should allow to specify (and to reason about) the system functions independently of their implementation that may be subject to timing and resource constraints. This gives the advantage that the functional correctness can be validated efficiently without taking into account the complex timing behavior of the implementation. 
When the system inputs (e.g., sampled data from sensors) are time-dependent, the system output is also time-dependent. 
In such cases, we need to reason about streams of reals (called here time streams), representing time points at which sensor data are sampled or outputs are written.
Time streams are simply another type of input and output streams for the system functions. 
In fact, time streams are generated by the system scheduler,  
and in turn used to sample the input and output streams computed by the system functions. 
\smallskip
}

{\bf Separation of computation and communication.}
A system model should allow to specify the components of the system for computations and the communication channels for data exchange separately, as not only independent units in the architecture but also in the semantics. We assume non-blocking data exchange, implemented by either asynchronous FIFO channels for buffering system inputs and outputs, or registers for storing sampled time-dependent data. 
The separation allows the system components to be specified as \emph{independent real-time tasks}, whose timing behaviours can be analyzed efficiently and locally. More importantly, as it is well-recognized in the theory of real-time scheduling \cite{stigge2014}, the underlying schedulability analysis for deployment will be greatly simplified compared with the case for dependent real-time tasks.
\smallskip

{\bf Updatability (avoidance of interference).}
The model of a system should allow for modifications by integrating new components for new system functions or replacing the existing components with refined ones, without changing the existing system functions determined by the original model.
The separation of computation and communication by asynchronous data exchange avoids inter-component interference when new components are integrated.
We require that new components may read but never write to the existing components via FIFOs or registers unless writing operations by the new components fulfill given requirements (specified using e.g. contracts \cite{DBLP:conf/fmics/GrafQGG18}), which is essential for future updates to preserve the original system functionality. Even though protocols may be needed to coordinate data exchange among the components (e.g. to avoid race conditions in register reading and writing), the components may operate autonomously or independently from each other even when some of them stopped functioning correctly. 
\smallskip

The rest of the paper is organized as follows: Section~\ref{sec:related} summarizes our contributions and related work. Section~\ref{sec:customer} presents the \customer{}
model, its informal semantics and the main theorem of this paper, stating the desired properties of \customer.
Section~\ref{sec:denotational} develops a fixed-point semantics for the model, establishing formal proofs for the main theorem.
Section~\ref{sec:analysis} presents open verification problems on the model, to be addressed in future work.
Section~\ref{sec:conclusion} concludes the paper. \vspace{-1.5mm} %
\section{Contributions and Related Work}\label{sec:related} 
One of the main challenges in embedded real-time systems design is to ensure that the resulting system has deterministic input-output and predictable timing behavior (typically with deterministic input-to-output latency or known time bounds)
even when multiple system functions are integrated and co-execute on a platform with limited resources.
The deterministic semantics allows model-in-the-loop simulation using successful tools like Simulink/Stateflow to simulate and verify the complete system behavior. Over the past decades, numerous approaches to address this challenge have been devised by research communities in hardware, software, control, and communication. Several, including the \textit{synchronous approach}, embodied by the languages Esterel, Lustre, and Signal \cite{nicolas}, and the time-triggered paradigm promoted by Kopetz \cite{kopetz}, ensure deterministic behavior by scheduling computation and communication among components at pre-determined time points.
This results in highly reliable and predictable systems, but severely restricts the possibility to modify or update systems after deployment. The reason is that new components must fit exactly into the already determined time schedules, and components may perturb each others' timing via shared resources. 
In recent years, dynamic updates of real-time systems after deployment have attracted increasing interest. A  model-based approach to the design and dynamic updates for cyber-physical systems is proposed in \cite{yiicfem17}. The work of  \cite{DBLP:conf/ahs/DorflingerAFMMS19} demonstrates that autonomous systems in operation can be updated 
through contract negotiation and run-time enforcement of contracts.
\smallskip

{\bf Contributions.}
We present a semantic model for real-time systems which on the one hand, ensures the deterministic input-output and predictable timing behaviors of a system, and on the other hand supports incremental updates after deployment without re-designing the whole system. In this model, a real-time system is described as a network of software components connected by communication channels. We provide a simple but expressive description language named \customer
to formalize such networks where each component is designed to compute a collection of functions over data streams and the communication channels can be of two types: FIFO queues for buffering inputs and outputs, and registers for sampling time-dependent data from sources such as sensors or streams that are written and read at different rates.
Components are further specified as real-time tasks to enforce that they read inputs, compute, and write outputs at time points satisfying certain time constraints.
A fixed-point semantics is developed for the model, showing that it enjoys two desired properties: (1) such a network of real-time software components computes a set of functions, each one defined from a set of (timed) input streams to a unique (timed) output stream.
(2) The network can be modified by integrating new components for adding new system functions or replacing existing components by refined ones (e.g. for better performance or security patches) without re-designing the whole system or changing the original system functions. \smallskip

{\bf Related Work.} 
An example of a time-triggered language developed for real-time systems is Giotto \cite{henzinger2003}. 
A Giotto program is a set of periodic tasks that communicate through ports.
Giotto implements the synchronous semantics, preserving timing determinism and also value-determinism (but not determinism over sequences of values i.e., streams as in our model) by restricting to periodic tasks where reading from and writing to ports is fixed and performed at deterministic time points.
It does not allow asynchronous communication via FIFO channels as \customer. 
This limits the possibility of updating a system in operation.
A more recent work addressing the quasi-synchronous semantics of \cite{Caspi2000} is presented in \cite{baudart:tel-01507595}. The work also proposes to use multiple periodic tasks to implement the synchronous semantics on parallel and distributed architectures. It remains in the category of synchronous approaches to real-time programming without addressing issues related to dynamic updates.
\customer{} can be viewed as a timed extension of Kahn Process Networks (KPN) \cite{kahn1974}.
In the literature, there have been various extensions to KPN. A special case of KPNs is dataflow process networks (DPN) \cite{lee1995}. A DPN is a general dataflow model where each process is specified as repeated firings of a node.
A node becomes enabled for execution according to a set of \emph{firing rules}. However, no time constraints are specified in the firing rules.
An implementation of KPN with bounded-size buffers is proposed in \cite{cohen2006}. In this work,  a composition approach preserving the Kahn semantics is  presented for components whose production and consumption rate are the same in the long run. 
The work, however, is confined within the synchronous programming model.
Related to the communication channels of KPN, a time-aware implementation of C, called \emph{Timed C}, has been proposed in \cite{natarajan2018}. In Timed C, a program consists of a set of tasks  communicating through two types of channels: \emph{FIFO} and \emph{Latest Value (LV)}. Analogous to KPN, reading from FIFO is blocking while writing is non-blocking. In contrast, reading and writing of LV channels are non-blocking. This communication model is similar to \customer. However, while Timed C is a general programming language without guaranteed determinism, we focus on both functional and timing determinism, and study these properties in a well-defined formal semantics.
A standardized software architecture for automotive domain is developed by AUTOSAR \cite{autosar}. Based on this, an application is organized as a collection of software components which perform data communication through a sender/receiver model. Data is processed by a receiver using a \emph{queue} or a \emph{last-is-best}  policy. Our model can be thought of as a specialization of this approach  which has a formal and deterministic semantics.
Due to the known fact that AUTOSAR is only a reference model for automotive software architecture with various implementations and without a  formal semantics, any formal proof is impossible. \vspace{-1.5mm}
\section{The \customer{} Model}
\label{sec:customer}

In this section, we present \customer{} based on Kahn Process Networks (KPN) \cite{kahn1974}.
A KPN is an abstract model of a parallel system consisting of a collection of processes connected by FIFO queues for data exchange. We view real-time systems as such a network where the computations as well as the respective input and outputs of the processes must meet given time constraints. 
Our model can be viewed as a timed version of KPN whose nodes are extended with timing constraints 
and edges with registers for sampling time-dependent inputs in addition to FIFO queues.

As KPN, \customer{} is essentially a simple description language to formalize system models. In this section, we present the main primitives and informal semantics of \customer. A formal fixed-point semantics is given in Section~\ref{sec:denotational}.

\subsection{Preliminaries on Kahn Process Networks}
\label{sec:customer:KPN}
\old{This subsection reviews the original} Here, we recall the notion of KPN and its main properties. A KPN is a set of stand-alone processes, called nodes, which communicate through a set of \emph{FIFO} channels. A node accesses channels through two operations: \textbf{read} and \textbf{write}.

\begin{definition}[KPN]
\label{def:KPN}
A Kahn Process Network $\mathcal{N}$  
is a set of processes, called nodes, %
and a set of FIFO queues, called \emph{channels}.
Nodes behave according to the following rules. \vspace{-1mm}
\begin{itemize}
    \item Each node computes a tuple of functions. For a set of input sequences, each of the functions defines a unique output sequence.   
    A node may not access the state/data of the other nodes.
      \item Channels are of potentially unbounded capacity. At most one node is allowed to read from / write to each channel. However, a node may copy an output to multiple channels read by multiple readers.
       
    \item {\bf read} from a channel is blocking, that is, \old{reading from an empty channel,} the node is blocked until \old{there is enough data} all data required for the next step is available, implying that a node cannot examine the emptiness of the channels;  {\bf write} to a channel is non-blocking. %
\end{itemize}
\end{definition}

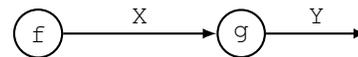
\begin{figure}[H]
\centering 

    \begin{tikzpicture} [scale=0.9]
    \node (o2) at (5,2) {}; 
    \node[circle,draw,thick,minimum size=.6cm] (f) at (0,2) { \code{f}}; 
    \node[circle,draw,thick,minimum size=.6cm] (g) at (3,2) { \code{g}}; 

    \draw[thick,->,>=latex] (f) to node[above]  { \code{X} } (g) ;
    \draw[thick,->,>=latex] (g) to node[above] { \code{Y} } (o2) ;
    \end{tikzpicture}

\caption{Structure of the program in Listing~\ref{list:kpn}.}
\label{f:kpnexample}
\end{figure}

 A node of a KPN can be implemented by a set of local variables and a procedure, repeated indefinitely.
The procedure may be specified in any conventional programming language, e.g., C.
\begin{example}
\label{ex:kpnexample}
An example of a KPN program is shown in Listing~\ref{list:kpn}. Nodes are defined by the \verb|process| keyword. 
The procedure executed by a node is written in a \verb|Repeat| block.
The structure of this program is depicted in  Fig.~\ref{f:kpnexample}, where arrows represent FIFO channels. %
\end{example} 

\noindent 

\begin{lstlisting}[caption={A sample KPN program.},emph={process,read,write},emphstyle=\textbf,language=c,label=list:kpn]
 process f(int out V) {
	 Repeat {		 write 1 on V;	 }
 }
 process g(int in U; int threshold; int out V) {
	 int count = 0;       // local variable
	 Repeat {
		 read(U);						// read from a channel
		 count = count + 1;
		 if count == threshold  
			 write 1 on V; 		// write to a channel
			 count = 0;	
	 }
 }
 int channel X, Y;
 f(X) || g(X, 5, Y);    // concurrent execution
\end{lstlisting}

\smallskip  

A KPN can be seen as a parallel program computing a set of functions from a set of input streams to a set of output streams defined by the least fixed point obtained by computing node functions in an arbitrary order~\cite{kahn1974}. Streams refer to complete histories of elements seen on some FIFO or output, and they are formally defined in Section~4.

The most important property of KPNs is their \emph{determinism}. This holds under any sufficiently fair scheduler, i.e., schedulers which do not postpone a process indefinitely. 

\begin{theorem}[Functional Determinism of KPN \cite{kahn1974}]
\label{th:KPN}
Given a set of input streams, the set of output streams computed by a KPN is unique. %
\end{theorem}

Theorem~\ref{th:KPN} indicates that implementation aspects, such as execution order, scheduling, and platform speed do not affect the functional behavior of a system implementing a KPN model.

\subsection{Timed KPN (TKPN)} \label{sec:customer:TKPN}

The order- and speed-independent functional determinism of KPN leads to a natural formalization of a timed version of KPN. Real-time systems are modelled as a KPN where each node is executed according to a real-time task model \cite{stigge2015}, specifying a release pattern which is a sequence of time points over the time line, and a deadline for each release.

\begin{definition}[TKPN]
\label{def:TKPN}
The timed version denoted $\mathcal{N}_T$ of a KPN $\mathcal{N}$ is obtained by associating with each node $\Nod$ of $\mathcal{N}$ %
a  release pattern and a positive integer,  called \emph{deadline}. %

\end{definition}

The release pattern of a TKPN node can be described using the well-established real-time task models \cite{stigge2015}, such as periodic tasks~\cite{liu1973}, 
generalized multiframe~\cite{baruah1999}, %
DAG~\cite{baruah2012}, or 
DRT~\cite{stigge2011} and
timed automata~\cite{fersman2007} as long as they are deterministic.

Note that each node of a TKPN computes a tuple of functions, one for each output channel. %
If different functions have different time constraints, different deadlines may be assigned to the respective output channels.

Note also that in Definition~\ref{def:TKPN}, the internal structure and resource requirement for the nodes of a TKPN and the scheduling algorithm to be adopted in the implementation are left open. Only the time constraints (i.e. the release patterns and deadlines for the executions of nodes) are specified. 

Informally, the operational behavior of a node in $\mathcal{N}_T$ is defined as follows. 
When the node is released,  and if all needed inputs are available, it computes and delivers the resulting outputs, if any, within the given deadline.
To achieve timing determinism, the inputs of a node are read at release time, outputs are \old{computed and} delivered at the deadline. This read-execute-write approach is similar to the \emph{implicit} communication model of AUTOSAR~\cite{autosarrte}. A formal semantics of TKPN is presented in Section~\ref{sec:denotational}.

Because a TKPN is also a KPN, and the execution rates assigned to nodes 
only restrict more explicitly the computation order of eligible nodes, the behaviour of a TKPN enjoys the desired functional determinism, which follows directly from Theorem~\ref{th:KPN}. 
Furthermore, it enjoys also the timing determinism {as declared later in Theorem~\ref{th:main}.} %
\vspace{-1mm}

\subsection{\customer: TKPN with Further Extensions} \label{sec:customer:registers}
In this section, we present our complete model. For this, we first extend TKPN with a new type of channel, called registers for sampling time-dependent data. Next, we augment the model with merge nodes for non-blocking reading of data from different sources. 

\paragraph{TKPN with register-channels.}
In real-time applications, inputs may be produced by the  physical environment, and hence, the corresponding value may be time-dependent. In Cyber-Physical Systems, such inputs come typically from sensors sensing physical phenomena. The system usually does not need all data produced by a sensor but only the latest value. Additionally, the refresh rate of the sensor is not necessarily compliant with the execution rate of the node(s) reading the sensor. In this case, using a FIFO may lead to memory overflow or blocked computation (in case the FIFO is empty). In such situations, it is useful to have a communication channel which keeps only the most recently written value. We extend TKPN with such channels, called \emph{register}.

The operations to access registers are syntactically the same as the ones to access FIFOs. 
We adopt the ``last-is-best'' semantics of \cite{autosar}. \textbf{write} to a register over-writes the current value. \textbf{read} from a register is non-blocking. When both \textbf{read} and \textbf{write} occur at the same time, the current value is
updated by \textbf{write} before it can be \textbf{read}. 

\begin{figure}[H]
  \centering
    \begin{tikzpicture}  [scale=0.9]
    \node (o2) at (5,2) {}; 
    \node[circle,draw,thick,minimum size=.6cm] (f) at (1,3) {\code{f}}; 
    \node[circle,draw,thick,minimum size=.6cm] (h) at (1,1) {\code{h}}; 
    \node[circle,draw,thick,minimum size=.6cm] (g) at (3,2) {\code{g}}; 

    \draw[thick,->,>=latex] (f) to node[above]  { \code{X} } (g) ;
    \draw[thick,->,>=latex,dashed] (h) to node[below]  { \code{Z} } (g) ;
    \draw[thick,->,>=latex] (g) to node[above] { \code{Y} } (o2) ;
    \end{tikzpicture}

    \caption{The structure of the program in Listing~\ref{list:tkpn}. Dashed arrow indicates a \emph{register}. }
    \label{f:tkpnexample}
          \end{figure}
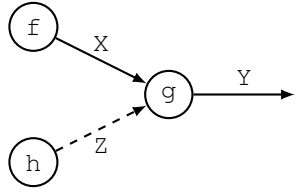

\begin{example}
\label{ex:customer}
Listing~\ref{list:tkpn} shows the program in Listing~\ref{list:kpn} extended with a register and time constraints. The program structure is illustrated by Fig.~\ref{f:tkpnexample}, where  FIFO channels are represented by solid-line arrows, and registers by dashed arrows.  %
\end{example}

In this example, using a register instead of a FIFO to carry the \emph{threshold} values has the advantage to (1) always use the most recent value available (the currently valid one), and (2) guarantee absence of buffer over- or underflow independently of the speed at which threshold values are produced and read.

\noindent 
\begin{minipage}{\linewidth}
      \centering
      \begin{minipage}{0.95\linewidth}
\begin{lstlisting}[caption={A sample program in Extended TKPN.}, emph={process,read,write},emphstyle=\textbf,language=c,label=list:tkpn]
 process f(int out V) { ... }         // unchanged
 
 process h(int out V) {
	 Repeat {		 write 6 on V;	 }
 }
 
 process g(int in U; int in C; int out V) {
	 int count = 0;
	 int threshold;
	 Repeat {
		 read(U);							// reading (from FIFO)
		 count = count + 1;
		 threshold = read(C);	// reading (from register)
		 if count >= threshold then 
			 write 1 on V; 				
			 count = 0;	
	 } 
 }
 // Instantiating and connecting the components.
 int channel FIFO X, Y;
 int channel register Z = 5;	 // initial value
 f.timings = periodic(10, 10); // period=deadline=10
 g.timings = periodic(10, 10);
 h.timings = periodic(10, 10);
 f(X) || h(Z) || g(X, Z, Y);
\end{lstlisting}
      \end{minipage}
  \end{minipage}

\paragraph{TKPN with merge-nodes.}
We further extend the model with a special type of nodes called \emph{merge}. A merge node has at least two FIFO inputs. At each activation, the node reads all data %
available in its input FIFOs. The output, which is written to a FIFO, is all the data read from the first input FIFO, followed by the ones from the second input FIFO, and so on. That is, a prioritized timed merge where all data items arriving during a release period are considered to "have arrived at the same time".

The motivation for defining this type is that in reactive systems, there are (external) events, e.g., requests for the same services from different sources, to which the system must react. No input event is allowed to be missed. Suppose now that events come from distinct producers so that distinct FIFO's are needed. Assume that the node should do a computation whenever there is some data in any input FIFO. This is beyond the expressiveness of KPN. Our time-dependent merge-function solves this very typical problem, common in real-time applications.

\begin{definition}[\customer: Extended TKPN]\label{def:customer:TKPN}
TKPN extended with registers and merge is TKPN where some channels can be a register instead of a FIFO and some nodes may be a merge-function. We call this extension of TKPN \customer.
\end{definition}

The formal semantics for \customer{} developed in Section~\ref{sec:denotational} shows that determinism is preserved if we see them as functions of the input data streams {\em and} their arrival times.

\begin{theorem}[Functional and Timing Determinism of \customer]
\label{th:main}
Given a set of input data streams and the arrival times of data items in the FIFO's (or registers), the set of output data streams computed and the time points at which the data items are inserted into the output FIFO's (or registers) by an extended TKPN is unique.
\end{theorem}

\noindent The result follows from Proposition~\ref{prop:registers} of Section~\ref{sec:denotational}.  %

\subsection{Design and Update with \customer %
} 
A model-based approach can be sketched as follows\footnote{Addressing the different steps in details, including specification, modelling, verification and compilation is not in the scope of this paper.}. 
First, to build a new 
system, a set of system functions to be implemented must be specified in terms of functional and timing requirements on their inputs and outputs as well as the respective end-to-end latency %
(see Section~\ref{sec:analysis}).
A \customer{} model may be constructed, verified to satisfy the given requirements and compiled into code executable on the target platform to compute these functions. 
Prior to any update over the life cycle of the system, its original \customer{} model may be extended (i.e., updated) by connecting the outputs of the existing components (KPN nodes) to the new ones.
Additionally, existing components may be replaced also by new ones fulfilling given requirements.
Thanks to the independence of reading from/writing to channels, the added (or updated) system functions will not interfere with the existing ones. Thanks also to the deterministic semantics, it can be verified based on the updated model that the resulting system will satisfy the functional and timing requirements. 
Further, it must be verified that the platform is able to provide enough resources to meet the resource requirements of the new components by schedulability analysis and analysis of memory usage (see Section~\ref{sec:analysis}). %
If all verification steps are successful, the new components can be deployed (or installed). Otherwise, the update is rejected.

\vspace{-1.5mm} %
\section{Fixed-Point Semantics}
\label{sec:denotational}

Here, we present a formal semantics for \customer. We first recall definitions used in \cite{kahn1974} to prove the order and time independent determinism of KPNs of Definition~\ref{def:KPN}.
We introduce a notion of timed stream to define the semantics of extended TKPNs of Definition~\ref{def:customer:TKPN} and to
prove the main theorem of Section~\ref{sec:customer:registers}. %

\subsection{Preliminaries on the Semantics of KPN%
} \label{sec:semantics:KPN}

First, we recall some basic notations from~\cite{kahn1974}. The function $\Fun$ associated with a node of a KPN is represented as a function from a set of input streams \old{, denoted $I$,} to a set of output streams\old{, denoted $O$}. More precisely, $\Fun$ represents a tuple of functions, one for each output.\smallskip

We now formally define streams and functions.
We consider streams on a generic domain $\Val$ which may be instantiated by any data domain. To ensure generality, we consider the time domain to be reals \Time.

\begin{definition}[Streams, time streams and timed streams\ignore{I limit to infinite streams so that the defs are simpler}]
\label{d:streams}
Let the stream domain $\Streams$ be the set of \ignore{finite and} infinite sequences in $\Val^{\infty}$. The domain of time streams $\Timed$ is the set of \ignore{finite and} infinite sequences of time points in $\Time^{\infty}$ with (not necessarily strictly) increasing time points which diverge\ignore{if the sequence is infinite, and terminate in time point $\infty$ if the sequence is finite}\footnote{That is, we assume time streams to be non Zeno.}.

The domain of {\em timed streams} $\Streams \times \Timed$ are \ignore{finite and} infinite sequences in $(\Val \times \Time)^{\infty}$ such that
every timed stream can be denoted as  $\Str\times\TStr$ for two appropriate streams. \ignore{of the same length or as one stream of pairs}

We use $\pref$ to stand for the standard prefix order on sequences,  $\lambda$ for the empty sequence, and  "$\,\conc\,$" for concatenation. 
\end{definition}

Note that a time stream may be regarded as a particular case of a data stream.
\old{Note also that $(\Streams,\pref,\lambda)$, $(\Timed,\pref,\lambda)$ and $(\Streams \times \Timed,\pref,\lambda)$ define each one a complete partial order.}
As in \cite{kahn1974}, functions $\Fun$ are built from the following basic functions on streams.

\begin{definition}[Functions on streams] \label{def:kahn-functions}
Consider the following functions from $\Streams^k$ to \Streams: \vspace{-1.5mm}
\begin{enumerate}
\item Data transformations: lift any $k$-ary data transformation function $\fun: \Val^k \mapsto \Val$ to a stream transformation function $\Streams^k\mapsto \Streams$, with the same name: $\fun(a_1\conc\Str_1,\ ...\ a_k\conc\Str_k)$ = $\fun(a_1,\ ...\ a_k) \conc \fun(\Str_1,\ ...\ \Str_k)$. \smallskip
\item Standard order preserving stream manipulating functions  "first", "remainder" and "append" (which we rarely use explicitly): 
$\first(a\conc\Str) = a$;
$\rem(a\conc\Str) = \Str$ (skips the first element of a stream);
$\mem(\Str,i_0\conc init) = i_0 \conc \Str$ (adds an initial element to the left by pushing the input stream to the right).  
\end{enumerate} 
\end{definition}

\begin{example}[Illustrating Example] 
Consider node $g$ of Fig.~\ref{f:kpnexample} \old{(Example~\ref{ex:kpnexample})} with input $X$ and output $Y$. %
We give the equations for all output streams using functions of Definition~\ref{def:kahn-functions}. Node $g$ has a local variable $count$ which gives rise to 2 streams: $c_M$, the stored values used as input of $g$, and $c$, the values produced by $g$.  \code{threshold} is a constant parameter $thsh$. Note that $g$ is independent of the actual data read, it just consumes a data item at each iteration. We obtain the following fixed-point equations:\vspace{-1.5mm}
\begin{itemize}
\item $Y$  = $g_Y(X, c_M, thsh)$ 
\item $c$\  = $g_{c}(X, c_M, thsh)$ 
\item $c_M$ = $\mem(0,c)$ 
    \quad \quad
    where
\item $g_{c}(x\conc X, c\conc c_M)$ = \  
[if $(c+1 < thsh)$ then $(c+1)$ else $0$] $\conc\ g_c(X,c_M)$ 
\item $g_Y(x\conc X, c\conc c_M)$ = \  
[if $(c+1 < thsh)$ then $\lambda$ else $1$] $\conc\ g_Y(X,c_M)$  
\end{itemize}
\end{example}
A typical function applies some transformation to the first elements of the input streams, produces an output (or alternatively produces nothing), and is applied recursively to proceed with the remainder of the streams. 
But a function may in each recursion step read zero or more elements from some input streams, and write zero or more elements to its output, as long as inputs are read in a FIFO order, the number of elements to be read is deterministically defined, and there is some "progress". \vspace{-1.5mm}

\subsection{Semantics of Timed KPN} \label{sec:semantics:TKPN}

We define the semantics of a timed node with a release pattern and a deadline. 
\New{In order to do so, we show that we can extend each function $\Fun$ on data streams (the semantic function for one of the functions of a KPN) to a function $\Fun_{\delta}$ on timed streams, such that (1) $\Fun_\delta$ defines a pair of streams consisting of the data stream defined by $\Fun$, and the stream of time points at which data elements are written.
(2) $\Fun_{\delta}$ is an ordinary Kahn function if time streams are just considered as particular data streams.
(3) the time extension corresponds to the intuition of release pattern $P$ and the output delay $\delta$ of Definition~\ref{def:TKPN}.
}
\New{We now state the proposition, the remainder of the subsection is dedicated to its proof}.
\ignore{First, we define the functions on timed streams which allow us to define this extension in a way that $\Fun$ is not changed by timing.}

\begin{proposition}\label{prop:TKPN}
The semantics of a TKPN is a deterministic mapping from timed input streams to timed output streams %
defined by a set of functions $\Fun_{\delta}$.
\old{the time point associated with each data item is the arrival time in its FIFO. This can be done by defining for every function $\Fun$ a function $\Fun_{\delta}$.
for every function $\Fun$ a function $\Fun_{\delta}$ whose output is a timed stream $\Str\times\TStr$ where $\Str$ is the output defined by $\Fun$ and $\TStr$ is deterministically defined on the timed input streams. }
\end{proposition}

\begin{figure}[t]
\center{
\scalebox{0.65}{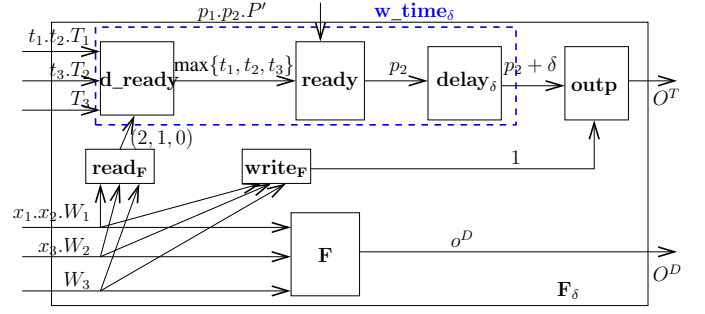}
}
\caption{Graphical representation of $F_\delta$}
\label{fig:F_delay}
\end{figure}

\New{We prove this proposition by constructing a function $\Fun_{\delta}$ for any data stream transformation $\Fun$. Fig.~\ref{fig:F_delay} illustrates $\Fun_{\delta}$ for a function $\Fun$ with 3 input streams. The data output is the one produced by $\Fun$. The time points associated with data elements are meant to represent the time point at which the data is inserted into the FIFO. Suppose that it holds for the input streams. This motivates our method for computing the time stamps of output data. It works as follows: (1) calculate the maximal time point associated with the data elements read by $\Fun$ (the time when all required data are in the FIFO), (2) calculate the actual "release" or "ready" time which is the release point of $P$ just after data is ready. (3) the output writing time point is obtained by adding $\delta$ and (4) this time point is output if and only if \Fun\ outputs a data item at this step. First, we introduce the necessary auxiliary functions:}\vspace{-1.5mm}

\begin{enumerate}
\item 
``data-ready'' function $\dready$: takes as input $k$ time streams $\TStr_i$, one for each input data stream of $\Fun$, and for each stream the number of elements to be read, and outputs the time when all data is ready: $\dready(({t_1}_1 ... {t_1}_{\ell_1}\conc\TStr_1,\ ...\ {t_k}_1 ... {t_k}_{\ell_k}\conc\TStr_k),$ $({\ell_1}\conc L_1,\ ...\ {\ell_k}\conc L_k))$ = $max\{{t_1}_{\ell_1},\ ... \ {t_k}_{\ell_k}\}$ $ \conc$ $\dready((\TStr_1,$ $ ...\ \TStr_k),(L_1, ...\ L_k))$. 

In each time stream, ${t_i}_{\ell_i}$ is the time stamp of the most recent data, hence $max\{{t_1}_{\ell_1},\ ... \ {t_k}_{\ell_k}\}$ is \old{the largest time stamp associated with the data to be read, that is,} the global {\em ready time} of the input. Some $\ell_i$  may be zero, meaning that no data is read from the corresponding stream, and there is no ${t_i}_{\ell_i}$ contributing to the maximum.

\item 
``ready'' function $\ready$: given time streams $P$ and $R$ (representing respectively a "release pattern" and "data ready times"), the stream of ``ready times'' for computation, the time points from which the deadline runs, is obtained by eliminating elements of $P$ at which no data is ready: 
$\ready(p\, \conc\, P', r\, \conc\, R')$ =  if $(p \geq r)$ then $p\, \conc\, \ready(P',R')$ else $\ready(P', r\conc R')$.

\item 
``delay'' function $\delay$: increases all time points of a time stream
 by a "delay" $\delta$. For a stream $\delta \geq 0$ (i. e. element wise $\geq 0$),
  $\delay( \TStr, \delta) = \TStr + \delta$. If $\delta$ is a constant, we note $\delay_{\delta}(\TStr)$ the corresponding function with a single argument. 

\item 
Given a $k$-ary function $\Fun(X^D_1,\ ... \ X^D_k)$ on data streams, we denote \\ $\readF(X^D_1,\ ... \ X^D_k)$ and $\writeF(X^D_1,\ ... \ X^D_k)$ two functions which read the same inputs as $\Fun$. \readF\ outputs a $k$-tuple of integers indicating the number of elements that \Fun\ reads at each computation step, and \writeF\ outputs the number of elements that \Fun\ outputs at each computation step.
For any function \Fun, these functions can be obtained by ``code analysis''\footnote{By ``code'', we mean the definition of \Fun\ in terms of the basic functions of Definition~\ref{def:kahn-functions}.}. \ignore{Actually defining such an analysis is out of the scope of this paper. }

\item 
 \outp: given a time stream and a stream of numbers in $[0 ... 1]$, it outputs the time point if the number is 1: $\outp(t\conc T, c \conc C)$ = if $(c==1)$ then $t \conc \outp(T,C)$ else $\outp(T,C)$. It is easy to extend to the case $c>1$, where the time point is written more than once. We use this function in a context where $c$ represents the number of elements written by a function $\Fun$, meaning that \outp\ guarantees that the number of time points written matches the number of elements written by $\Fun$.
\end{enumerate}

\ignore{(was old)For a $k$-ary function $\Fun(X^D_1,\ ... \ X^D_k)$, we define now a corresponding function $\Fun_\delta$ taking into account a release pattern $P$.
\Fun\ may read several elements at a time from each input stream, and we introduce some notations to ease writing: 
$X^D_i = s^D_i \conc W^D_i$ where $X^D_i$ are the current input streams, $s^D_i$ the sequences to be read, and $W^D_i$ the streams remaining to be transformed.

\noindent Then, suppose $\Fun(X^D_1,\ ... \ X^D_k)$ = $f(s^D_1,\ ...\ s^D_k) \conc \Fun(W^D_1,$ $ ...\ W^D_k)$ where $f$ is the data transformation of the step. This means that \Fun\ produces one data item. Alternatively, one may have $\Fun(X^D_1,\ ... \ X^D_k)$ = $\Fun(W^D_1,$ $ ...\ W^D_k)$, that is, \Fun\  produces no data item for $(s^D_1,\ ...\ s^D_k)$.%
\footnote{We make this restriction for the sake of conciseness. Functions writing more than one data item can be handled easily.} 
}

Now we can define $\Fun_\delta$ by composing the previously defined functions as suggested by Fig.~\ref{fig:F_delay}, where function $\deadl_{\delta}$ is represented by a blue dashed box.\ignore{, and all other involved functions are those defined above.}

\begin{itemize}
\item[6. ] $\Fun_\delta((X_1,$ $ ...\ X_k),P) $ = $(\Fun(X^D_1,\ ...$ $ X^D_k),$ $\outp(\deadl_{\delta}((X^T_1,\ ... \ X^T_k),\readF(X^D_1,$ $... X^D_k), P),$ $\writeF(X^D_1, ... X^D_k))$ where 
\item[7. ]$\deadl_{\delta}((X^T_1,$ $ ... \ X^T_k),$ $(\ell_1,\ ...\ \ell_k), P)$ = $\delay_\delta$ $(\ready(P,$ $\dready((X^T_1,$ $ ... $ $X^T_k)$, $(\ell_1,\ ...\ \ell_k))))$. \vspace{-1mm}
\end{itemize}

\New{To sum up, $\Fun_\delta$ has the promised characteristics: (1) it defines a pair of streams consisting of the data stream defined by $\Fun$, and a stream of time points at which data elements are written. This proofs that $\Fun$ is preserved.} \old{Time points are computed by function $\deadl_\delta$. At each iteration step of \Fun\ at which no (resp. one) data is produced, \outp\ guarantees that also no (resp. one) time point is produced.}
\ignore{Finally, note that the $k$-tuple of integers  $(\ell_1,\ ...\ \ell_k)$ which are used by $\dready$ to compute the data ready times are provided by the call of $\readF(X^D_1, ... X^D_k)$.}
\New{(2) It is an an ordinary Kahn function if time streams are considered as particular data streams. This proves timing determinism based on \cite{kahn1974}.}
\old{For functions defined as a composition of functions of Definition~\ref{def:kahn-functions}\ignore{and such that each recursion step contains at least one remainder function (at least one element is consumed at every step)}, determinism of the system functions follows from \cite{kahn1974}. Any $\Fun_\delta$ is such a function. $\Fun_\delta$ also preserves $\Fun$, and the time stream produced corresponds to the intuition of Definition~\ref{def:TKPN}.} 
\New{(3) It produces time streams according to the intuition of Definition~\ref{def:TKPN}.
}
This completes the proof. %

Section~\ref{sec:customer:TKPN} makes the choice that deadlines run from the first release point after all data is ready. The definitions here correspond to this choice. But the definition of $\Fun_\delta$ can easily be adapted to alternative choices.

According to Definition~\ref{def:customer:TKPN}, $\delta$ represents an exact output write delay, that is, the data is actually written into its FIFO at the time point defined by $\Fun_{\delta}$. Note that exactly the same definition may be used to represent a best or worst case execution time or a "latest due time", for example. 

\begin{example}[Illustrating example, continued]
Consider again function $g$ of Fig.~\ref{f:kpnexample}. Now, $g$ is executed periodically with period \code{period} (noted \code{p}), and has a output delay \code{deadline} (noted \code{dl}). 
As functions on data remain untouched, we only need to add equations defining the time streams associated with $Y$, $c$ and $c_M$ using the previous definitions. As we know that at each step exactly one data is read from each input, the definition of timed streams is slightly simplified. Note that function $g_Y$ may or may not produce output, whereas $g_c$ always produces output:\vspace{-1mm}
\begin{itemize}
\item $Y^T$ = $\outp(\deadl_{\code{dl}}((X^T, c_M^T),(1,1),$  
$\code{p}),$ $\writes_{g_Y}(X^D, c_M^D))$,  
\item $c^T$ = $\outp(\deadl_{\code{dl}}((X^T, c_M^T),(1,1),$ 
$\code{p}),$ $1)$,\ and  
\item $c_M^T$ =  $c^T$.              
\end{itemize}
For function $M$ representing the variable \code{count}, we consider that once the variable is written (defined by time stream $c^T$), it is available immediately for the next step, and no additional delay is added. \vspace{-1.5mm}
\end{example}

\subsection{Semantics of \customer%
} \label{sec:semantics:registers}

We now provide the semantic underpinning for the full model \customer{} (Extended TKPN) by defining also  "register" and "merge" as transformers of timed streams. As motivated in Section~\ref{sec:customer:registers}, such functions depend on the data that is put in the FIFO (or a register) at the release time points of the function. At the semantic level, this means that they read all the (not yet treated) data items with time stamp up to some time point -- the release time of the function -- and compute some output depending on these data items.

As motivated in Section 3, a merge-node is a timed stream transformer merging incoming timed streams into a timed stream -- a total order -- according to the time points at which data items inserted in the respective FIFO's. We view registers also as a timed stream transformer transforming the incoming timed stream to a timed stream containing data items (and also their arrival times) according to the register-reading ratio, i.e., the release pattern of the target node. 

\New{We define the semantics of Extended TKPN and prove its determinism.
In order to do so, we show that we can define $k$-ary functions $\sens$ and $\merge$ from timed streams to timed streams
(1) which correspond to the intuition of the concepts "register" and "merge" introduced in Section~\ref{sec:customer:registers}, and 
(2) which are ordinary Kahn functions if time streams are considered as ordinary data streams
}
\ignore{at the semantic level, this requires to represent a register as a transformation of the stream produced by the source component into the value stream read by the target component.}

\New{We first state the proposition that, together with Proposition~\ref{prop:TKPN}, guarantees  Theorem~\ref{th:main}, and the remainder of the subsection is dedicated to its proof.}

\begin{proposition}[Extended TKPN]\label{prop:registers}
The semantics of an Extended TKPN is a deterministic mapping from timed input streams to timed output streams.
\old{In a TKPN with register inputs \New{and merge functions} all (timed) output streams are uniquely defined by their timed input streams. Furthermore, the resulting streams correspond to the intuition of a ``register'' \New{and of ``timed merge''} as defined in Section~\ref{sec:customer:registers}.} \ignore{read on a register input are exactly those defined by function $\sens$. }
\end{proposition}

\New{We prove this proposition by constructing the above mentioned functions $\sens$ and $\merge$ which together with the previously defined functions $\Fun_{\delta}$ define the semantics of Extended TKPN.}
\old{First, we define registers and merge as Kahn functions. The transformation ``register read" (denoted \sens) has two arguments, the timed data stream $(\Str,\TStr)$ written to the register, and a time stream $\sampTStr$representing the time points up to which the data are to be read at every activation. The output is a timed data stream representing the timed data actually read.}\smallskip

$\sens$ has two arguments, the timed data stream $(\Str,\TStr)$ written to the register, and a time stream $\sampTStr$ representing the time points up to which the data are to be read at every activation. The output is a timed data stream representing the (timed) data actually read. For the sake of simplicity, suppose that \TStr\ starts at time 0 and $\sampTStr$ at a time $\geq 0$.

\begin{itemize}
\item[1. ]
Define
$\sens( (\Str,\TStr),\ \sampTStr)$ 
inductively:\\
$\sens((s_1,t_1)\conc (s_2,t_2)\conc (\Str,\TStr), t\conc \sampTStr)$ = if $t_2 < t$ then  $\sens((s_2,t_2)\conc (\Str,\TStr), t\conc \sampTStr)$ else  $(s_1,t_1)\conc \sens((s_1,t_1)\conc(s_2,t_2)\conc (\Str,\TStr), \sampTStr)$. \\
We may denote $\sens$ as a pair of functions $(\sens^D,\sens^T)$. \vspace{-1mm}
\end{itemize}

\noindent The if-clause represents the case where the first element $(s_1,t_1)$ is an ''overwritten data'' to be skipped, and the else-clause the case where $(s_1,t_1)$ is the newest data with a time stamp prior to $t$.
Note that in this second case, $(s_1,t_1)$ is not ``consumed'' but reread at the next iteration, so as to guarantee that for any time point $t'$ of the remaining $\sampTStr$, there is at least one data older than $t'$. 
Elements that are written into the "register" but "overwritten" before the next time point of reading are filtered out, and elements that are to be read more than once are written several times. This represents the functionality of a ``register'' read at the time points $\sampTStr$. 
\smallskip

\New{$\merge$ has as arguments (at least) three arguments, timed data streams $(\Str_i,\TStr_i)$ representing the FIFOs to be merged, and a time stream $\sampTStr$ representing the time points up to which the data are to be read and merged at each activation. The output is a timed data stream representing the merged FIFO and their data arrival times.}

\begin{itemize}
\item[2. ]
Define
$\merge( (\Str_1,\TStr_1),\ (\Str_2,\TStr_2),\ \sampTStr)$ 
inductively:\\
$\merge( (d_1,t_1)\conc(\Str_1,\TStr_1),\ (d_2,t_2)\conc(\Str_2,\TStr_2),\ t\conc\sampTStr)$ =\\
 if $t_1<t$ \ignore{and $t_1\leq t_2$} then $(d_1,t)\conc \merge((\Str_1,\TStr_1),\ (d_2,t_2)\conc(\Str_2,\TStr_2),\ t\conc\sampTStr)$\\
 else if $t_2<t$ \ignore{and $t_1 < t_2$} then $(d_2,t)\conc \merge( (d_1,t_1)\conc(\Str_1,\TStr_1),\ (\Str_2,\TStr_2),\ t\conc\sampTStr)$\\
 else\footnote{that is, $t_1,t_2\geq t$ meaning no data to be merged up to $t$ is left} $\merge( (d_1,t_1)\conc(\Str_1,\TStr_1),\ (d_2,t_2)\conc(\Str_2,\TStr_2),\ \sampTStr)$\vspace{-1mm}
 \end{itemize}

\noindent \New{That is, data with time stamps inferior to $t$ are taken from the two input streams  by first moving the data items from the left queue, and then those from the right. This corresponds to the informal definition which does not order elements of one reading interval according to their arrival times but only according to the priority, and their time stamps are the "sample time points". Proceeding to the next time point of time stream $\sampTStr$ (if no relevant data item is left) corresponds to a new activation of the merge function.}\smallskip

\New{To sum up, the functions $\sens$ and $\merge$ have the required characteristics: (1) they produce the intuition of the concepts "register" and "merge" introduced in Section~\ref{sec:customer:registers}, and (2) they are Kahn functions. All functions of an Extended TKPN can be composed from functions of the form $\Fun_\delta$, by \sens\ or by \merge. This guarantees their determinism.
This completes the proof. %
}

\old{As $\sens$ \New{and $\merge$} are composed from functions of Definition~\ref{def:kahn-functions}, they are deterministic.  }
\old{The $\merge$ function orders data of different reading intervals according to their time stamps, and data from the same reading interval according to a global priority. If $\sampTStr$ represents the activation times of the merge function (and time stamps represent arrival time in the FIFO), this corresponds to the "merge of FIFO contents" described in Section~3.
$\sens$ reads the most recent data written before the time points determined by the second argument. If the second argument represents reading times (the release time of the node), this corresponds exactly to a register. \ignore{If they are older time-points, it rather corresponds to a "delayed register"} This completes the proof. %
}

\ignore{But, as one may notice from Definition~\ref{def:register},  determining the value of $\sens$ at time point $t_0$, may require to read also a time point {\em later than} $t_0$. 
Indeed, we do not use time just as a particular data stream, but we want to compute the required value {\em at} time $t_0$. Only if at time $t_0$ exactly the elements of \Str\ with a time value $t\leq t_0$ have been written to the register,
then function \sens\ defines exactly the stream defined by ``reading the register at time points of \TStr''.
} \smallskip

\New{Function $\merge$ can be implemented if the data items present in the FIFO at the activation time points can be deterministically defined. If data items can be written at sufficiently precisely defined time points, this is the case.}
The guarantee that data items must be present at a certain time point, also allows to detect when there is "no data item available in a given interval", which may be either just normal or mean that a run-time error has occurred and trigger some exception handler.
 \smallskip

Note also that the condition on deterministic writing times cannot be relaxed without losing the guarantee of determinism if \sens\ is implemented by a simple register (that is memory). \New{If time points associated with data streams represent "latest'' writing times for example, determinism can nevertheless be preserved by using more than just one memory: one for the data to be ready at the next release time, and one or more memories for release time points further in the future}. Similarly, to achieve the timed merge function one would need to group data according to the activation period of their "latest writing times". This is an adaptation of Caspi's protocol defined in \cite{CaspiMP2001} to our framework. Note however, that in both cases it requires to time stamp data explicitly at implementation level.

\ignore{An advantage of both alternatives is that they allow to relax strict timing determinism of read and write which is a strong constraint.}
\ignore{Note however, that the determinism of data read from FIFOs and of functions depending only on FIFO inputs is not affected by timing non determinism.}

\begin{example}[Illustrating example, continued]\label{ex:fixedpointTKPN}
Again, consider function $g$ (now Fig.~\ref{f:tkpnexample} of Example~\ref{ex:customer}). Now, \code{threshold} is defined by an input $Z$, a register. %
This may an important effect, on both time and data. \vspace{-1.5mm}
\begin{itemize}
    \item 
    As a register holds a valid data at any time, $Z^T$ does not influence \dready, and time streams do again not depend on $Z^T$. 
    \item 
    The data streams of $Y$ and $c$ have now 3 input streams and are computed by replacing parameter $thsh$ by $\sens^D(Z,\ready(\code{p},\dready(X^T,c^T_M)))$.  \vspace{-1.5mm}
\end{itemize}
\end{example} %
\section{Analysis Problems}
\label{sec:analysis}

To update a system by adding a new system function, its model (in \customer) should be modified to integrate the new function. To enable that {the modified model} can be compiled into a program executable on a platform with limited resources, it must be verified to meet given timing and resource constraints. 
Thanks to the deterministic semantics of \customer,  properties verified on a \customer{} model will be preserved by the execution of the compiled code on a platform satisfying the resource assumptions adopted in the verification process. 

There are two principle timing and resource constraints: (1) the memory requirements must be bounded and (2) system functions including the new one must satisfy the end-to-end latency constraints \cite{abdullah2019}.
In general, these verification problems are undecidable. However, with proper assumptions and restrictions, there are efficient solutions for practical purposes \cite{krcal2006,abdullah2019}. First, the required memory of a system in operation depends on the buffer size required by the FIFOs, which can be specified as follows.

\begin{definition}[Required buffer size (\rbs)]
\label{def:buffsize}
Assume that the data written to and read from a FIFO buffer are specified, respectively, by timed streams $(a_1,t_1)(a_2,t_2)...$ and $(a_1,t_1')(a_2,t_2')...$.
Also, let
    $\omega(t)= \max\{i| t_i \leq t \}$  and %
    $\gamma(t)= \max\{i| t_i' < t \} $.
The FIFO's \emph{required buffer size} (\rbs) is defined as 
$\max\{ \omega(t) - \gamma(t) | t \geq 0\}$. %
\end{definition}
In words, $\omega(t)$ is the total number of items written to a FIFO up to (including) time $t$, and $\gamma(t)$ is the total number of items read from the FIFO  strictly before $t$. Based on this, the \rbs\ of a FIFO  denotes the maximum number  of items which may simultaneously exist in the queue. 
Indeed, computing \rbs\ in a \emph{process network}  has been shown undecidable in the general case~\cite{buck1993}.
Despite this, the measure is computable for special settings. %
For instance, if for each node, the number of produced and consumed items is fixed in all firings, as is the case in synchronous data flow  (SDF)~\cite{lee1987}, the problem has efficient solutions~\cite{lee1987}. 
Further, for those KPNs in which data producing/consuming pattern of the nodes is periodic (except for a bounded initial time), it is shown that the required capacity for a FIFO is bounded if and only if writing and reading rates are  \emph{asymptotically} the same~\cite{cohen2006}. 

The \rbs\ of a \customer{} model depends on the release pattern of the nodes,  the pattern by which input data arrives, and also the data consumption pattern. According to these factors, a variety of instances of the problem of computing (a bound on) \rbs\ %
can be defined, which  we leave for future work. Here, we just provide some initial observations.

A fairly direct consequence of Definition~\ref{def:buffsize} is that 
the \rbs\ of a FIFO in a \customer{} model %
is bounded if and only if a constant $c$ exists for which:
 $\forall t \geq 0 :  \omega(t) - \gamma(t) \leq c $.
Based on this, we conjecture that:
\emph{The \rbs\ of a FIFO is bounded if and only if reading and writing rates are asymptotically the same, i.e., 
$\lim_{t \rightarrow \infty} \{\omega(t) / \gamma(t) \} = 1$. %
} 
\smallskip

The other measure to be analyzed is \emph{end-to-end latency}, which essentially reflects the responsiveness of a system. It is important that when the input changes, or for an event arriving in the input queue, the system provides a response (or react) within a bounded delay.  %
As an essential requirement in a real-time system~\cite{abdullah2019}, we define the end-to-end latency for each output channel of the system with respect to an input on which it depends.

\begin{definition}[Worst-case end-to-end latency, $\etoe(i)$] %
Consider {\color{black} a system function specified by} a $k$-ary function $\SysFun$ on streams. Let $(I_1, \ldots I_k)$ be a set of input timed streams for which $\SysFun(I_1, ... I_k) = (Y^D,Y^T)$. Consider now, for some $i$, $1\leq i \leq k$, a modified version of $I_i$, called $I'_i$, which is obtained by changing the $j$-th entry of $I_i$ from $(a,t)$ to $(b,t)$. Assume  $\SysFun(I_1, \ldots,I'_i,\ldots I_k) = (Y'^D,Y'^T)$. Let $j' = \min \{m | Y'^D_m \neq Y^D_m \mathrm{\ or\ } Y'^T_m \neq Y^T_m\}$, where  $X_m$ denotes the $m$-th element  of a stream $X$. We define 
 $delay(I_1, ... I_k,i,j,b) = Y'^T_{j'} - t$.
 Accordingly, we define
$ 
delay(I_1, ... I_k,i) = \max \{delay(I_1, ... I_k,i,j,b) | \forall j,b \}
$.
The worst-case end-to-end latency for the $i$-th input line is then
 \[
 \etoe(i) =  \max \{delay(I_1, ... I_k,i) | \forall (I_1, ... I_k) \}   %
\]  

\end{definition}

Computing end-to-end latency can be studied in terms of the release patterns of the nodes in a \customer{} model. 
For instance, for a set of periodic tasks communicating through a set of registers, which can be viewed as a special case of Extended TKPN, the problem has been explored in~\cite{kloda2020} which presents
a worst-case analysis method with exponential time complexity. Also, a polynomial-time approach is provided for computing an upper bound. Both methods are limited to task sets scheduled with a fixed-priority policy on a single processor. 
Investigating the problem in \customer{} for different release patterns and target platforms such as multi-core and distributed architectures is left to future work. \vspace{-1.5mm}

\section{Conclusions}\label{sec:conclusion}
This paper presents a deterministic timed model (\customer) to enable a model-based approach allowing for not only building deterministic real-time systems, but also verifying and updating them after deployment. 
\customer{} is a timed extension of Kahn's Process Network with (1) timing constraints on the execution of KPN nodes and (2) register-channel and merge-node to deal with time-dependent data and functions. \customer{} is proven to preserve functional and timing determinism. More precisely,
given a set of input data streams and the corresponding arrival times of data items in the input channels of a \customer{} model, the set of output data streams computed by its network nodes, and the time points at which the data items are inserted into its output channels is unique.

To further develop a programming (or coordination) language based on \customer, and a compiler for such a language, several challenging verification problems (see Section \ref{sec:analysis}) must be solved. First, a program (or a \customer{} model) must be analyzed to ensure that its memory requirement is bounded and meet the platform limitations. Second, the end-to-end latency for system functions computed must satisfy given timing requirements. Future work includes also a symbolic semantics for the model allowing uncertainty in the implementation e.g. when data exchange through reading from and writing to channels may occur over time intervals.  
\bibliographystyle{splncs04}
\bibliography{ref}

\end{document}